\documentclass[letterpaper]{article} 
\usepackage{aaai24}  
\usepackage{times}  
\usepackage{helvet}  
\usepackage{courier}  
\usepackage[hyphens]{url}  
\usepackage{graphicx} 
\usepackage{natbib}  
\usepackage{caption} 
\usepackage{algorithm}
\usepackage{algorithmic}

\usepackage{newfloat}
\usepackage{listings}
\usepackage{threeparttable} 
\DeclareCaptionStyle{ruled}{labelfont=normalfont,labelsep=colon,strut=off} 
\floatstyle{ruled}
\newfloat{listing}{tb}{lst}{}
\floatname{listing}{Listing}
\title{Exploring the Impact of ChatGPT on Student Interactions in Computer-Supported Collaborative Learning}
\author{
    Han Kyul Kim\textsuperscript{\rm 1}, Shriniwas Nayak\textsuperscript{\rm 1}, Aleyeh Roknaldin\textsuperscript{\rm 1}, \\ Xiaoci Zhang\textsuperscript{\rm 1}, Marlon Twyman\textsuperscript{\rm 1}, Stephen Lu\textsuperscript{\rm 1} 
}
\affiliations{
    \textsuperscript{\rm 1}University of Southern California\\

    \{hankyulk, spnayak, roknaldi, xiaocizh, marlontw, sclu\}@usc.edu
}

\usepackage{bibentry}

\begin{document}

\maketitle

\begin{abstract}
The growing popularity of generative AI, particularly ChatGPT, has sparked both enthusiasm and caution among practitioners and researchers in education. To effectively harness the full potential of ChatGPT in educational contexts, it is crucial to analyze its impact and suitability for different educational purposes. This paper takes an initial step in exploring the applicability of ChatGPT in a computer-supported collaborative learning (CSCL) environment. Using statistical analysis, we validate the shifts in student interactions during an asynchronous group brainstorming session by introducing ChatGPT as an instantaneous question-answering agent.
\end{abstract}

\section{Introduction}
A generative AI such as ChatGPT brings a paradigm shift in education settings with its potential usage in providing personalized learning or virtual assistance to instructors \citep{Baidoo-Anu2023EduInGenAI, Lo2023ChatGPTSurvey, Qadir2023EnggEduInChatGPT}. However, concerns related to hallucination \citep{ji2023survey, alkaissi2023artificial} and academic integrity \citep{mcmurtrie2023teaching, rudolph2023chatgpt} have been raised, casting doubt on the broader applicability of generative AI in education.

The issues of factual accuracy and fair assessment from generative AI are crucial when the primary objective of learning activities is to acquire accurate factual information. Yet, it's essential to recognize that such an objective, in line with established education frameworks like Bloom’s taxonomy \citep{bloom1956handbook, kropp1966validation}, represents only a fraction of the learning objectives necessary for effective educational experiences. When learning activities are designed to empower students to analyze, evaluate, or create new ideas, as categorized in Bloom's taxonomy, fostering students' creativity becomes a core issue \citep{egan2017developing, al2018review}. For example, in brainstorming activities, the quantity or novelty of the ideas become crucial metrics to track \citep{fu2015design, hong2016understanding}.

In this paper, we present a unique application of ChatGPT within the context of asynchronous computer-supported collaborative learning (CSCL), in which the focal learning objective is to enhance students' creative thinking. Our contribution not only expands the application of generative AI in educational settings but also provides initial statistical analyses illustrating the changes in student interaction patterns resulting from integrating ChatGPT into a group brainstorming activity.

\section{Experiment setting \& methods}
\subsection{CSCL context}
Our application and analysis draw from student chat data gathered in a graduate-level product engineering course during the spring of 2023. Over the final two weeks of the semester, 12 cohorts, each comprising 4 to 5 students, were randomly generated. Each week, cohort members participated in asynchronous online group discussions for three days to generate a list of creative product design ideas aligned with specific customer requirements. At the end of each week, their ideas were presented to the entire class and reviewed by both their classmates and instructors.

Among 12 cohorts, 6 cohorts, labeled as \texttt{Type A}, were randomly selected to incorporate ChatGPT as an additional question-answering agent. Students in these cohorts were provided with the ability to tag ChatGPT in their messages. Subsequently, our internal system interfaced with ChatGPT's OpenAI API to retrieve and post its responses to the tagged messages on the group's chat. Conversely, the remaining 6 cohorts, referred to as \texttt{Type B}, were not provided with any option to directly interact with ChatGPT during their discussions. The specific number of students in each cohort and comprehensive statistics on the number of messages in each cohort are provided in Tables \ref{tab:StudentCount} $\sim$ \ref{tab:typebstat} in the Supplementary materials.

\subsection{Evaluation methods}
To evaluate the impact of ChatGPT on student interactions during group brainstorming activities, we employed the Collaborative Learning Conversation Skills Taxonomy (CLCST) \citep{soller2001supporting}. It classifies the interaction-related skills commonly used by students in CSCL environments. Two experts familiar with CLCST manually identified messages in the discussions that exhibited the students' active learning, as defined in \citet{soller2001supporting} as follows:

\begin{itemize}
\item Request: ask for help/advice in solving the problem or in understanding a team-mate's comment.
\item Inform: direct or advance the conversation by
providing information or advice.
\item Motivate: provide positive feedback and reinforcement.
\end{itemize}

To ensure annotation consistency, we computed Cohen’s kappa \citep{cohen1960coefficient} between the two annotators, resulting in a consistency score of 93.2\%. This high value indicates almost perfect agreement between the two annotators \citep{mchugh2012interrater}.

For all our statistical analyses, we conducted repeated ANOVA, incorporating the inclusion of ChatGPT in a cohort and the week of the brainstorming session as two independent covariates. To account for the variation in the number of messages and students in each cohort, we use the proportions of active learning messages within each cohort as a dependent variable. Upon obtaining significant results from the F-test, we employed the Tukey method \citep{tukey1949comparing} for subsequent multiple pairwise comparisons, allowing us to mitigate group-wise variation.

\section{Results}
When comparing the level of active learning in overall interactions within \texttt{Type A} and \texttt{Type B} cohorts, we did not observe any statistical significance (Table \ref{tab:anova_overall1} in the Supplementary materials). While the impact of ChatGPT was not evident across all student interactions, this result does not preclude the possibility of statistical significance within different types of student interactions. As ChatGPT was exclusively included in \texttt{Type A}, we can categorize the student interactions within \texttt{Type A} into two distinct types: student-student interactions and student-ChatGPT interactions.

\subsection{Student-student interaction}
As students from cohorts in \texttt{Type A} were required to tag ChatGPT in their messages to induce its response, all of their messages without the tag were regarded as intended for other students. For \texttt{Type B}, where ChatGPT was not available as a question-answering agent, all of their interactions were categorized as student-student interactions. 

\begin{table}[!h]
\centering
\caption{ANOVA results on the ratio of active learning within student-student interactions, indicating statistically significant group-wise variation resulting from introducing ChatGPT}
\begin{tabular}{lrrr}
\hline
Source of Variation & SS & df & F \\
\hline
ChatGPT & 0.307 & 1 & $\textbf{4.909}^{*}$ \\
Week & 0.031 & 1 & 0.501 \\
ChatGPT $\times$ Week & 0.080 & 1 & 1.285  \\
Residual & 1.253 & 20 & \\
\hline
\end{tabular}
\begin{tablenotes}
\centering
\item[] * p $<$ 0.05, ** p $<$ 0.01, *** p $<$ 0.001
\end{tablenotes}
\label{tab:anova_ss1}
\end{table}

\begin{table}[!htb]
\centering
\caption{Impact of ChatGPT on the ratio of active learning in student-student interaction based on Tukey's method, indicating a higher ratio of active learning in \texttt{Type B} cohorts}
\begin{tabular}{lcc}
\hline
Comparison & Difference & Adj. p-value \\
\hline
With ChatGPT vs. & 0.226 & \textbf{0.037} \\
Without ChatGPT & & \\
\hline
\end{tabular}
\label{tab:tukey_result1}
\end{table}

Based on the statistical analysis from Tables \ref{tab:anova_ss1} and \ref{tab:tukey_result1}, cohorts in \texttt{Type B} exhibited a 22.6 percentage point higher ratio of active learning in student-student interactions. This finding implies that even though there was no major change in the overall student interaction, the internal dynamics of student interactions, particularly the level of active learning, were significantly altered by incorporating ChatGPT.

\subsection{Student-ChatGPT interaction}

When analyzing student-ChatGPT interactions, we exclusively focus on cohorts in \texttt{Type A}, comparing the levels of active learning in messages directed towards ChatGPT and those addressed to other students within the same cohort.

\begin{table}[!h]
\centering
\caption{ANOVA results on the ratio of active learning between student-ChatGPT and student-student interactions, indicating statistically significant group-wise variation between the types of interactions}
\begin{tabular}{lrrr}
\hline
Source of Variation & SS & df & F \\
\hline
To ChatGPT & 1.722 & 1 & $\textbf{24.765}^{***}$ \\
Week & 0.000 & 1 & 0.000  \\
ChatGPT $\times$ Week & 0.008 & 1 & 0.11  \\
Residual & 1.391 & 20 & \\
\hline
\end{tabular}
\begin{tablenotes}
\centering
\item[] * p $<$ 0.05, ** p $<$ 0.01, *** p $<$ 0.001
\end{tablenotes}
\label{tab:anova_sc1}
\end{table}

\begin{table}[!h]
\centering
\caption{Impact of ChatGPT on the ratio of active learning in student-ChatGPT interaction based on Tukey's method, indicating a higher level of active learning between student interactions directed to ChatGPT}
\begin{tabular}{lcc}
\hline
Comparison & Difference & Adj. p-value \\
\hline
To ChatGPT vs. & -0.5357 & \textbf{0.000} \\
To students & & \\
\hline
\end{tabular}
\label{tab:tukey_result2}
\end{table}

The statistical results presented in Tables \ref{tab:anova_sc1} and \ref{tab:tukey_result2} indicate a significant statistical difference in the level of active learning between student interactions directed towards ChatGPT and those addressed to other students. On average, student-ChatGPT messages exhibited a 53.57 percentage points higher level of active learning compared to student-student interactions. This result suggests that the active learning process within cohorts was becoming more bilateral between students and ChatGPT, even in the presence of other students in the cohort.

\section{Conclusion}
Our paper presents preliminary findings from an interesting application of ChatGPT in group brainstorming activities in the context of CSCL. Our statistical analysis quantitatively confirms that ChatGPT diminishes the level of active learning interactions between students, a crucial element for successful collaborative learning. This decrease is primarily attributed to a relatively higher proportion of active learning interactions occurring predominantly between individual students and ChatGPT.

While our initial exploration sheds light on this phenomenon, further analysis is imperative to unravel its underlying causes fully. Moreover, a deeper dive into ChatGPT's influence on the tangible outcomes of group brainstorming sessions remains an essential avenue for future investigation. It's worth noting that our experimental design did not explicitly account for the potential novelty effect of introducing ChatGPT, a factor we aim to address in our future work. Nevertheless, it is essential to continue exploring the potential of generative AI in assisting students in creative thinking-related tasks. It provides a promising educational setting to leverage generative AI's capability without being significantly limited by its hallucination issues.

\section{Supplementary materials}

\begin{table}[ht]
\centering
\caption{Number of students in each cohort}
\begin{tabular}{|c|c|c|}
\hline
         & \texttt{Type A} & \texttt{Type B} \\
\hline
Cohort \#1 & 5        & 4        \\
\hline
Cohort \#2 & 4        & 5        \\
\hline
Cohort \#3 & 5        & 4        \\
\hline
Cohort \#4 & 5        & 5        \\
\hline
Cohort \#5 & 5        & 5        \\
\hline
Cohort \#6 & 5        & 5        \\
\hline
\textbf{Total} & 29        & 28        \\
\hline
\end{tabular}
\label{tab:StudentCount}
\end{table}

\begin{table}[!h]
\caption{Summary of messages in \texttt{Type A}}
\begin{tabular}{|cccc|}
\hline
\multicolumn{4}{|c|}{\textbf{\texttt{Type A}}}                                                                 \\ \hline
\multicolumn{1}{|c|}{\textbf{\begin{tabular}[c]{@{}c@{}}Cohort\\ number\end{tabular}}} & \multicolumn{1}{c|}{\textbf{\begin{tabular}[c]{@{}c@{}}Total \\ number\\ of student \\ messages\end{tabular}}} & \multicolumn{1}{c|}{\textbf{\begin{tabular}[c]{@{}c@{}}Avg \\ number\\ of words per\\ message\end{tabular}}} & \textbf{\begin{tabular}[c]{@{}c@{}}Total \\ number\\ of ChatGPT\\ messages\end{tabular}} \\ \hline
\multicolumn{1}{|c|}{Cohort \#1}                                                       & \multicolumn{1}{c|}{35}                                                                                        & \multicolumn{1}{c|}{20.371}                                                                                  & 13                                                                                       \\ \hline
\multicolumn{1}{|c|}{Cohort \#2}                                                       & \multicolumn{1}{c|}{83}                                                                                        & \multicolumn{1}{c|}{13.41}                                                                                   & 28                                                                                       \\ \hline
\multicolumn{1}{|c|}{Cohort \#3}                                                       & \multicolumn{1}{c|}{92}                                                                                        & \multicolumn{1}{c|}{31.011}                                                                                  & 10                                                                                       \\ \hline
\multicolumn{1}{|c|}{Cohort \#4}                                                       & \multicolumn{1}{c|}{115}                                                                                       & \multicolumn{1}{c|}{23.191}                                                                                  & 34                                                                                       \\ \hline
\multicolumn{1}{|c|}{Cohort \#4}                                                       & \multicolumn{1}{c|}{38}                                                                                        & \multicolumn{1}{c|}{26.789}                                                                                  & 18                                                                                       \\ \hline
\multicolumn{1}{|c|}{Cohort \#5}                                                       & \multicolumn{1}{c|}{46}                                                                                        & \multicolumn{1}{c|}{24.0}                                                                                    & 28                                                                                       \\ \hline
\end{tabular}
\end{table}

\begin{table}[!h]
\caption{Summary of messages in \texttt{Type B}}
\centering
\begin{tabular}{|ccc|}
\hline
\multicolumn{3}{|c|}{\textbf{\texttt{Type B}}}                     \\ \hline
\multicolumn{1}{|c|}{\textbf{\begin{tabular}[c]{@{}c@{}}Cohort\\ number\end{tabular}}} & \multicolumn{1}{c|}{\textbf{\begin{tabular}[c]{@{}c@{}}Total \\ number\\ of student \\ messages\end{tabular}}} & \textbf{\begin{tabular}[c]{@{}c@{}}Avg \\ number\\ of words per\\ message\end{tabular}} \\ \hline
\multicolumn{1}{|c|}{Cohort \#1}                                                       & \multicolumn{1}{c|}{25}                                                                                        & 17.56                                                                                   \\ \hline
\multicolumn{1}{|c|}{Cohort \#2}                                                       & \multicolumn{1}{c|}{85}                                                                                        & 14.729                                                                                  \\ \hline
\multicolumn{1}{|c|}{Cohort \#3}                                                       & \multicolumn{1}{c|}{30}                                                                                        & 11.767                                                                                  \\ \hline
\multicolumn{1}{|c|}{Cohort \#4}                                                       & \multicolumn{1}{c|}{61}                                                                                        & 15.18                                                                                   \\ \hline
\multicolumn{1}{|c|}{Cohort \#4}                                                       & \multicolumn{1}{c|}{21}                                                                                        & 29.691                                                                                  \\ \hline
\multicolumn{1}{|c|}{Cohort \#5}                                                       & \multicolumn{1}{c|}{71}                                                                                        & 19.239                                                                                  \\ \hline
\end{tabular}
\label{tab:typebstat}
\end{table}

\begin{table}[!h]
\centering
\caption{ANOVA results on the level of active learning within all student interactions}
\begin{tabular}{lrrr}
\hline
Source of Variation & SS & df & F \\
\hline
ChatGPT & 0.012 & 1 & 0.202 \\
Week & 0.080 & 1 & 1.289  \\
ChatGPT $\times$ Week & 0.032 & 1 & 0.515  \\
Residual & 1.237 & 20 & \\
\hline
\end{tabular}
\label{tab:anova_overall1}
\end{table}

\bibliography{aaai24}

\end{document}